\documentclass[10pt]{iopart}
\usepackage{iopams}
\expandafter\let\csname equation*\endcsname\relax
\expandafter\let\csname endequation*\endcsname\relax
\usepackage{amsmath}
\usepackage{graphicx}
\usepackage{xcolor}
\usepackage{hyperref}
\usepackage{cite}

\begin{document}

\title[]{3D CO-TALIF distribution above a micro cavity discharge: A systematic approach for plasma catalysis}

\author{Henrik van Impel$^{1}$, Oliver Krettek$^{2}$, David Steuer$^{1}$, Volker Schulz-von der Gathen$^{2}$, Marc Böke$^{2}$ and Judith Golda$^{1}$}

\address{$^{1}$Plasma Interface Physics, Ruhr-University Bochum, D-44801 Bochum, Germany}
\address{$^{2}$Experimental Physics II: Physics of Reactive Plasmas, Ruhr-University Bochum, D-44801 Bochum, Germany}

\ead{Henrik.vanImpel@rub.de}
\vspace{10pt}
\begin{indented}
\item[]March 2026
\end{indented}

\begin{abstract}
We investigate a micro cavity plasma array (MCPA) reactor operated at atmospheric pressure, offering excellent diagnostic accessibility and flexible opportunities for catalyst integration. Dissociation processes, such as CO production from CO$_2$ diluted in helium, are studied. The diagnostic setup, combining TALIF with an ICCD camera, enables three-dimensional spatially resolved measurements of CO number densities above the CO-generating discharges. The measured distributions are compared to a basic three-dimensional diffusion model, showing good agreement and revealing the dominant transport mechanisms. Flow variation studies indicate that the gas flow inside the reactor follows a laminar, Poiseuille-like profile, while the transport behavior is consistent with literature values for the diffusion coefficient of CO, further validating the model. The high estimated local dissociation within the MCPA discharges (about 40\,\%) results in saturation of CO production under increasing voltage. Combined with complementary diagnostics developed for this discharge, including measurements of surface charges, electric fields, and atomic oxygen, this approach provides a suitable platform for systematic studies of plasma–catalyst interactions.

\end{abstract}
\submitto{\PSST}
\maketitle

\ioptwocol
%
%
%
%
%

\section{Introduction}
Environmental pollution and climate change caused by anthropogenic CO$_2$ emissions represent some of the most urgent global challenges of the 21st century \cite{intergovernmental_panel_on_climate_change_ipcc_climate_2023}. Mitigating these impacts requires the development of sustainable and energy-efficient technologies capable of converting environmentally harmful gases into value-added chemicals and fuels. In this context, plasma-assisted conversion processes -- especially exploiting renewable energy as power sources -- have gained increasing attention, as non-thermal plasmas enable chemical reactions far from thermodynamic equilibrium by selectively energizing electrons while maintaining comparatively low gas temperatures. This unique characteristic allows the activation of chemically stable molecules such as CO$_2$ under mild operating conditions, offering new pathways for sustainable gas conversion \cite{liu_non-thermal_1999}.

Among these approaches, the combination of non-thermal plasma with heterogeneous catalysts has emerged as a particularly promising strategy to enhance reaction efficiency, improve product selectivity, and potentially reduce overall energy consumption\cite{Van_Durme_2008, bogaerts_plasma_2020}. The interaction between plasma-generated reactive species and catalytic surfaces can lead to synergistic effects that cannot be achieved by either plasma or catalysis alone. Despite significant progress in recent years, a fundamental understanding of these plasma–catalyst interactions remains limited, and identifying true synergistic mechanisms continues to be a major scientific challenge, especially disentangling the interplay between modifying surface characteristics and the plasma.

One key difficulty arises from the experimental investigation of plasma–catalyst systems. The integration of catalytic materials into plasma reactors frequently restricts diagnostic accessibility. This limitation is particularly pronounced in commonly used packed-bed reactor configurations, where optical access is strongly restricted and achieving spatially resolved diagnostics becomes extremely challenging \cite{touitou_situ_2013}. At atmospheric pressure, these challenges are further amplified because stable plasma operation typically requires electrode gaps on the micrometer scale, making the application of advanced diagnostics particularly demanding. Consequently, the investigation of fundamental plasma parameters such as local particle number densities, electric field distributions, and mean electron energies requires discharge geometries that simultaneously allow catalyst integration and spatially resolved measurements.

To address these challenges, we investigate a micro cavity plasma array (MCPA) reactor operated at atmospheric pressure, which provides both excellent diagnostic accessibility and flexible opportunities for catalyst integration. The modular dielectric barrier discharge (DBD) device consists of a high-voltage-driven metal foil, a dielectric layer that can be spray-coated with a catalyst, and a grounded magnet. During operation, dielectric barrier discharges are ignited inside thousands of laser-cut µm-sized cavities within the metal foil, forming an array of spatially separated micro-discharges. This configuration enables highly reproducible and quasi-independent plasma operation within each cavity while maintaining macroscopic scalability. Combined with spatially resolved diagnostics, the MCPA provides an excellent platform for investigating plasma–catalyst interactions with high statistical relevance, allowing systematic variation of key parameters such as material properties, electric fields, and surface charge effects \cite{Dzikowski_2022, vanImpel_2024, Labenski_2024}. 

Previous investigations have demonstrated that the MCPA represents a well-characterized plasma source and exhibits highly efficient production of reactive species, particularly atomic oxygen, making it a suitable model system for studying plasma chemistry \cite{Steuer_2023,Steuer_2024}. Furthermore, recent studies have shown pronounced plasma–surface interactions within the device, highlighting its strong potential as a platform for systematic plasma-catalysis investigations \cite{vanImpel_2025}.
To extend the MCPA reactor from the oxygen model system to application-relevant dissociation processes such as CO$_2$ or other carbon containing molecules, CO production is studied in CO$_2$ diluted in helium. Since carbon monoxide is less reactive than atomic oxygen, different plasma dynamics are expected, making it a valuable probe and mode system for understanding gas-phase processes in the MCPA reactor.

Here, spatially resolved CO number densities are measured using two-photon absorption laser-induced fluorescence (CO-TALIF), a well-established diagnostic technique in plasma research \cite{damen_absolute_2019}. Due to its high spatial resolution, this method is particularly well suited for investigating plasma–surface interactions and for resolving local distributions of reaction products such as CO.
In this approach, CO molecules are excited from the electronic ground state via resonant two-photon excitation, and the resulting fluorescence emission in the Ångström band system is detected. This enables the reconstruction of three-dimensional CO number density distributions within the MCPA reactor.
The measured density distributions are compared with a basic three-dimensional diffusion model, which shows good agreement with the experimental data. 



\section{Diagnostics}
\subsection{TALIF of CO}

Two-photon absorption laser-induced fluorescence (TALIF) is a well-established technique for determining absolute particle densities. In this work, it is applied to measure absolute CO number densities. The excitation scheme employs the $X^1\Sigma^+(v=0) \rightarrow B^1\Sigma^+(v'=0)$ transition (two-photon equivalent wavelength 230\,nm, corresponding to 115\,nm single-photon excitation), while the resulting fluorescence in the Ångström band system ($B^{1}\Sigma^{+}(v'=0) \rightarrow A^{1}\Pi(v''=2)$) is detected around 520\,nm, as commonly used in CO-TALIF measurements \cite{linow_comparison_2000,di_rosa_two-photon_1999,di_teodoro_collisional_2000}. 

This scheme offers several advantages compared to alternative excitation pathways (e.g., $X^1\Sigma^{+}\rightarrow C^1\Sigma^{+}$) for absolute CO density measurements at atmospheric pressure with nanosecond laser systems. In particular, collisional quenching rates are well documented in the literature, and the natural lifetime of the excited state is relatively long (about 22\,ns) \cite{di_teodoro_collisional_2000}. Moreover, this excitation scheme has been successfully applied by Damen et al.\ in a low-pressure DC CO$_2$ glow discharge, yielding results consistent with FTIR measurements \cite{damen_absolute_2019}. Because radiation at 115\,nm is strongly absorbed in air and difficult to generate and transport, the excitation is realized via a two-photon process.

To convert the measured fluorescence signal into an absolute number density, a calibration with a known reference density $N_{\mathrm{ref}}$ is required. In the case of CO, the calibration can be directly conducted by CO admixtures to helium from a CO bottle. The CO density $N_{\mathrm{CO}}$ is determined from the measured signal strengths $S$ and $S_{\mathrm{ref}}$ (each normalized to the squared laser intensity) according to

\begin{equation}
N_{\mathrm{CO}} =
\frac{S}{S_{\mathrm{ref}}}
\cdot
\frac{Q + \tau_{\mathrm{nat}}^{-1}}{Q_{\mathrm{ref}} + \tau_{\mathrm{nat}}^{-1}}
\cdot
\frac{\Phi^{(2)}(T_{\mathrm{ref}},\lambda_{\mathrm{ref}})}{\Phi^{(2)}(T,\lambda)}
\cdot
N_{\mathrm{ref}},
\label{formula:density}
\end{equation}

where $\tau_{\mathrm{nat}}$ denotes the natural lifetime of the laser-excited state, and $Q$ and $Q_{\mathrm{ref}}$ represent the total collisional quenching rates under measurement and reference conditions, respectively. Moreover, the normalized spectral shape $\Phi^{(2)}$ of the two-photon excitation rate coefficient is included. The quenching rate could not be determined from the measured fluorescence decay curves because the laser pulse duration and the effective lifetime of the excited state are of the same order of magnitude. Therefore, literature values for the quenching rates were used \cite{di_teodoro_collisional_2000}. 

In the experiments, three main quenchers were considered: He, CO, and CO$_2$, which contribute to the total collisional quenching rate Q

\begin{equation}
Q = p\left(q_{\mathrm{He}}f_{\mathrm{He}} + q_{\mathrm{CO}}f_{\mathrm{CO}} + q_{\mathrm{CO}_2}f_{\mathrm{CO}_2}\right).
\end{equation}

$p$ is the total pressure and the $f_{\mathrm{He}}$, $f_{\mathrm{CO}}$ and $f_{\mathrm{CO}_2}$ are the different fractions in the gas composition. Here, helium has the lowest quenching rate ($q_{\mathrm{He}}=0.083$\,MHz/mbar), CO$_2$ the highest ($q_{\mathrm{CO}_2}=19.5$\,MHz/mbar), while CO lies in between ($q_{\mathrm{CO}}=6.1$\,MHz/mbar). In the present measurements, the exact fractions of CO and CO$_2$ are not known due to the ongoing dissociation process. Therefore, two limiting cases are considered: no CO$_2$ dissociation and complete dissociation. Additional species such as O$_2$, O, and O$_3$ may also be produced and could contribute to quenching. In the present analysis, however, atomic oxygen and ozone are neglected. In the limiting case of complete CO$_2$ dissociation, the maximum O$_2$ fraction would be $\frac{1}{2}f_{\mathrm{CO}_2}$. Since the quenching rate of molecular oxygen ($q_{\mathrm{O}_2}=13.5$\,MHz/mbar) lies between those of CO$_2$ and CO, this situation is effectively covered by the two limiting cases considered.

Because CO molecules populate different rotational levels in the ground state characterized by the rotational quantum number $J$, and assuming a Boltzmann distribution, a rotational temperature $T$ can be defined. To account for this effect, the normalized spectral shape of the two-photon excitation rate coefficient $k^{(2)}$ at the excitation wavelength $\lambda$ is introduced and defined as

\begin{equation}
\Phi^{(2)}(T,\lambda)=
\dfrac{k^{(2)}(T,\lambda)}
{\int k^{(2)}(T,\tilde{\lambda})\,\mathrm{d}\tilde{\lambda}}.
\label{eq:phi}
\end{equation}

Populations of vibrationally excited ground-state levels are neglected in the following, as their occupation at near-ambient temperatures is expected to be very small. Nevertheless, the extent of vibrational excitation should always be verified. Further details on the derivation of the density expression can be found in \cite{damen_absolute_2019}.

\section{Setup}
\subsection{MCPA setup} 
\begin{figure*}[htb]
    \centering
    \includegraphics[width=\textwidth]{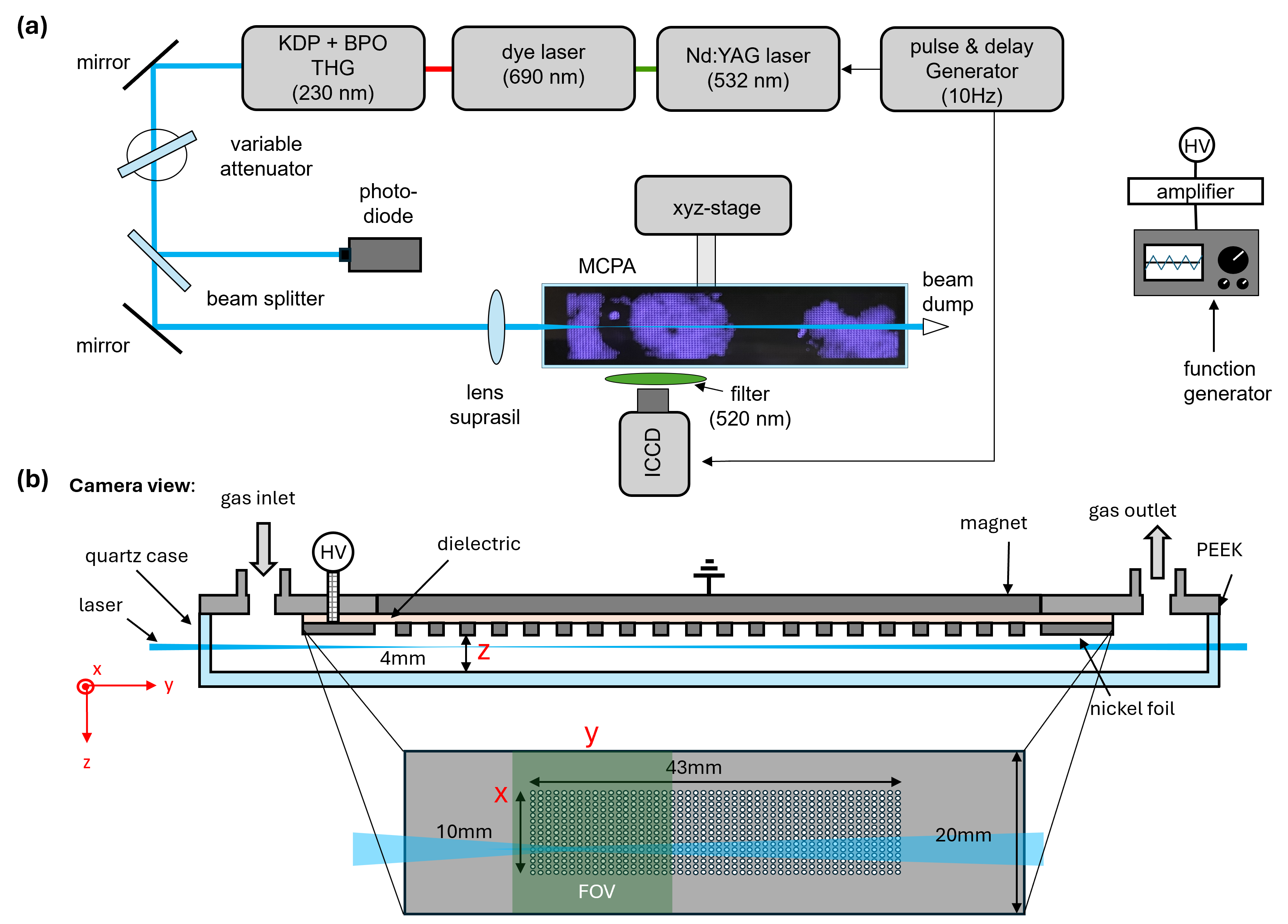}
    \caption{Laser setup for 3D TALIF measurements (a) and a close-up of the MCPA from the perspective of the fluorescence detecting ICCD camera (b).}
    \label{fig:TALIF_setup}
\end{figure*}

In this study, a micro cavity plasma array (MCPA) was used, whose basic design has been described in detail elsewhere \cite{Dzikowski_2020, Dzikowski_2022, Steuer_2023}. This dielectric barrier discharge (DBD) device consists of three main components, shown as a close-up in Fig.~\ref{fig:TALIF_setup}(b): \begin{itemize} \item A laser-cut nickel foil, 50\,µm thick, patterned with circular cavities of 200\,µm diameter spaced 150\,µm apart. The cavities form an array covering $10\times 42\,\mathrm{mm}^2$ (28x122 cavities). During operation, high voltage is applied to this foil. \item Beneath the nickel foil lies a 40\,µm thick dielectric foil (ZrO$_2$, relative permittivity $\epsilon_r\approx 27$), which forms the bottom of the cavities. \item Below both foils is a magnet that attracts the nickel foil and clamps the dielectric in place; it also serves as the counter electrode. \end{itemize} As a result, several thousand microdischarges ignite within these cavities. Further details on the overall discharge dynamics, including electric fields, surface charges, mean electron energies, and dissociation processes with oxygen as a model system, can be found in \cite{Dzikowski_2020, Kreuznacht_2021, Dzikowski_2022, Steuer_2023, vanImpel_2024, Labenski_2024, Steuer_2024, Steuer_2025, vanImpel_2025}. 

A function generator (Tektronix AFG 3021B) in combination with a high-voltage amplifier (Trek PZD700A) is used to apply a 15\,kHz bipolar triangular waveform with a voltage amplitude of approximately 600\,V to the nickel foil. An atmospheric pressure helium flow of 0.5–2\,slm with a CO$_2$ admixture (up to 1\,\%) is directed over the MCPA. To confine the flow while maintaining optical access, the MCPA is enclosed by a quartz cover. The plates are 3\,mm thick, and the cover is positioned 4\,mm below the nickel foil surface.

Increasing the CO$_2$ content in the MCPA reactor leads to incomplete ignition of the cavities. This effect is illustrated in Fig.~\ref{fig:TALIF_setup}(a), which shows a discharge image recorded at a gas flow of 1\,slm helium with an admixture of 4\,sccm CO$_2$. Under these conditions, only about 60\,\% of the cavities ignite. For the remaining cavities, breakdown is likely not achieved, possibly due to small geometric variations within the structure that lead to locally reduced electric fields insufficient to ignite the discharge at elevated CO$_2$ admixtures.

Furthermore, the ignited cavities appear to influence neighboring ones, as the image reveals distinct regions in which groups of cavities ignite collectively. Such collective dynamics and inter-cavity interactions have been discussed previously in \cite{Dzikowski_2020}. The partially ignited MCPA configuration, therefore, provides a suitable test case to demonstrate that CO-TALIF can resolve spatial structures and local variations on small length scales within the reactor. At the same time, the observed trends are not limited to this specific operating condition but can be generalized to MCPAs with a higher fraction of ignited cavities, owing to the spatially resolved nature of the diagnostic approach, as described in the following section
\subsection{Diagnostic setup}
The TALIF setup is shown in Fig.~\ref{fig:TALIF_setup}(a). As the system control unit, a digital pulse generator (Stanford Research Systems DG535) is used with a trigger frequency of 10\,Hz. It triggers a Nd:YAG laser (Continuum Powerlite 8000), which operates at 532\,nm (frequency-doubled), and pumps the dye laser system (Dye: Pyridin 1 in methanol). A laser wavelength in the range of 690\,nm is selected and then amplified. This wavelength is tunable with a step size of 1\,pm. Subsequently, the dye output is frequency-doubled and sum-frequency mixed to generate the desired 230\,nm for excitation of CO molecules. For this, a combination of a KD*P (potassium dideuterium phosphate) and a BBO (beta barium borate) crystal is used for frequency mixing. The generated laser beam has an energy of 190\,µJ and a pulse length of 6-7\,ns. The outgoing beam passes a beam splitter, where about 10\,\% of the beam is directed to a fast photodiode to measure the energy of the laser beam. The main beam is then focused by a Suprasil lens ($f=30\,\mathrm{cm}$) through the quartz cover into the MCPA reactor. After passing through the rear side of the quartz cover, it reaches a beam dump.

The laser-induced fluorescence signal within the MCPA reactor is detected by an ICCD camera (Andor iStar DH334T-18U-73) that is placed at a right angle to the laser beam in the horizontal plane (compare Fig.~\ref{fig:TALIF_setup}(b)). For focusing, a lens is used (Nikon ED AF MICRO NIKKOR 200mm 1:4D). The resulting pixel-to-pixel resolution is about 18\,µm. With a detector size of 1024x1024 pixel a field of view (FOV) of 18.4\,mm in both dimensions is possible. An optical filter (center wavelength 520\,nm, 10\,nm FWHM) is mounted directly in front of the camera. This filter ensures detection of the Ångstrom bands of the excited CO molecules and therefore reduces stray light, especially that generated by reflections of the laser beam within the quartz cover. The camera is triggered synchronously with the laser and then internally gated with a width of 100\,ns to further reduce noise. The gate width was chosen generously compared to the lifetime of the excited CO state at atmospheric pressure (a few ns), so that the camera trigger can be set a few ns before the incoming laser pulse to ensure that the entire signal is captured despite jitter.

The reactor is mounted upside down as described in \cite{Steuer_2024}. To enable the measurement of 3D profiles of the density distribution within the reactor, it is mounted on a system of three orthogonal linear stages (Zaber). The reactor is connected to a gas system consisting of three mass flow controllers (MKS MFC GE50A, 2\,slm, 20\,sccm, 5\,sccm) connected to three gas bottles: a helium bottle as carrier gas, a CO$_2$ bottle and a customized calibration bottle filled with 99\,\% helium and 1\,\% CO. \

\subsection{Evaluation of the TALIF signals}
The experimental setup produces two-dimensional images of the CO-TALIF signal, recorded with an integration time of 60\,s at each measurement position. The evaluation of these data represents a compromise between maximizing the signal-to-noise ratio and preserving sufficient spatial resolution. Here, the TALIF signal is defined as signal strength divided by the squared laser intensity.

The laser beam exhibits a Gaussian intensity profile with a full width at half maximum (FWHM) of approximately 190\,µm. To improve the signal-to-noise ratio, the TALIF signal was averaged along the beam profile in the z-direction (distance from the discharge) over a range of $\pm50$\,µm around the beam center. Within this interval, the laser intensity decreases by approximately 15\,\% relative to the peak value. Consequently, all measurements represent an average over a spatial extent of about 100\,µm, resulting in a reduced spatial resolution in the z-direction.

\begin{figure}
    \centering
    \includegraphics[width=\linewidth]{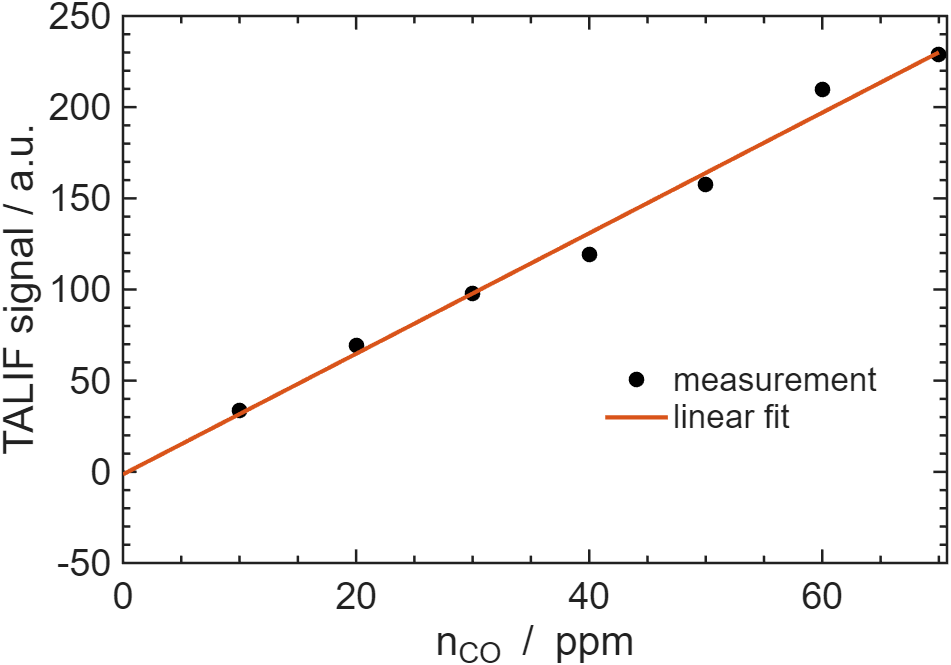}
\caption{Calibration measurement showing the TALIF signal as a function of the admixed CO calibration gas concentration. The orange line represents a linear fit to the data.}
    \label{fig:signalCali}
\end{figure}

To determine absolute CO number densities, a calibration measurement was performed. For this purpose, the spatially averaged TALIF signal recorded by the camera was measured while systematically varying the CO concentration by admixing a calibration gas with known CO content. The resulting calibration curve (Fig.~\ref{fig:signalCali}) exhibits an almost perfectly linear dependence, as expected for TALIF under the applied conditions. Importantly, the linear fit closely intersects the origin, indicating a high signal-to-noise ratio and negligible background contribution.
Following the signal averaging procedure described above, CO number densities were calculated using Eq.~(\ref{formula:density}). For each three-dimensional measurement position, a corresponding calibration was applied, and the CO density was determined on a pixel-by-pixel basis.

    \begin{figure}
    \centering
    \includegraphics[width=\linewidth]{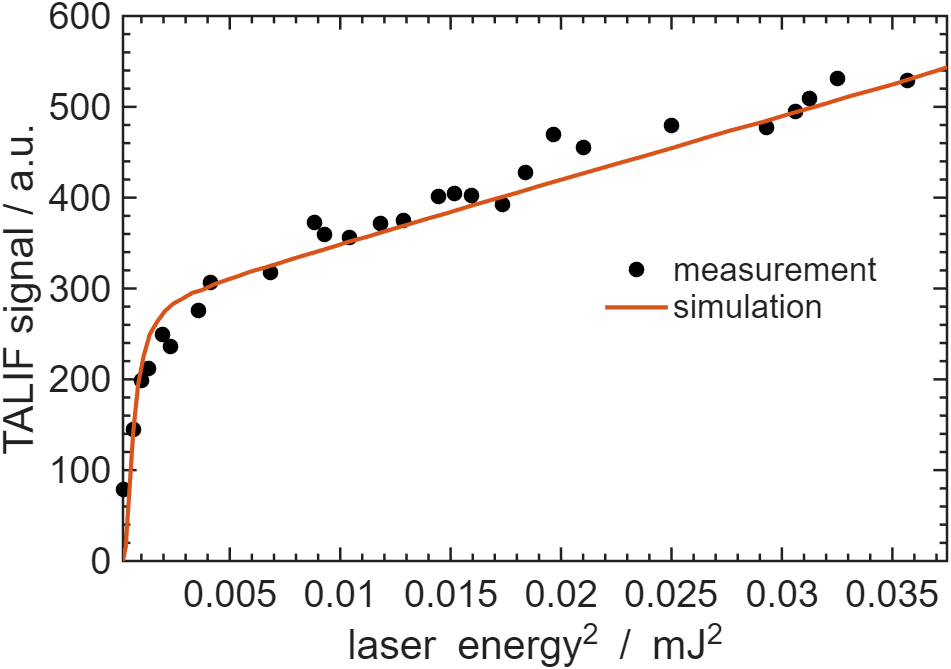}
\caption{Measured TALIF signal as a function of the squared laser pulse energy in a calibration measurement with 40\,ppm CO added to the gas flow. The orange curve shows a simulated signal assuming a Gaussian distribution of the laser pulse energy ($\sigma = 0.4$) and the presence of a detection threshold in the camera system.}
    \label{fig:signalVSpow}
\end{figure}    
In TALIF measurements, different excitation regimes can occur depending on the applied laser pulse energy. Since the excitation probability in two-photon absorption scales with the square of the laser intensity, the fluorescence signal is expected to follow a quadratic dependence on laser energy in the unsaturated regime. This quadratic regime is generally desired, as it provides the highest sensitivity to changes in species number density while avoiding saturation effects and intensity-induced spectral broadening (power broadening).

To verify the excitation regime, the laser pulse energy was systematically varied while simultaneously recording the corresponding TALIF signal. The resulting dependence is shown in Fig.~\ref{fig:signalVSpow}. At low laser energies, a steep increase of the signal is observed, which transitions into a region exhibiting an approximately quadratic scaling with reduced slope. No indication of saturation is observed within the accessible energy range. Consequently, measurements were performed at the highest available laser energy in order to maximize signal intensity while remaining within the non-saturated excitation regime.

The pronounced signal reduction at low laser energies is attributed to the single-shot detection limit of the ICCD camera. Assuming that the laser pulse energy follows a normal distribution with a standard deviation of approximately $\sigma = 0.4$ around the mean pulse energy (an effect that becomes particularly relevant at low energies) and introducing a detection threshold for the camera system, the experimentally observed curve in Fig.~\ref{fig:signalVSpow} can be reproduced (orange curve). This artifact occurs when a fraction of the individual laser shots fall below the detection threshold, resulting in an apparent deviation from the expected quadratic scaling. This effect should be taken into account when interpreting measurements at low laser energies, but will not be discussed further within the scope of this paper.

\subsection{Wavelength scan}

In order to reduce experimental complexity, measurements were performed at a single excitation wavelength rather than recording a full excitation spectrum at every spatial position. To justify this approach, it must be ensured that plasma operation does not significantly modify the rotational population distribution of ground-state CO molecules ($X^1\Sigma^+(v=0)$), which would otherwise affect the TALIF excitation efficiency in Eq.~(\ref{eq:phi}). Consequently, any difference in the rotational distribution between plasma-on and reference conditions must be taken into account when computing absolute densities from the reference measurement.

\begin{figure}
    \centering
    \includegraphics[width=\linewidth]{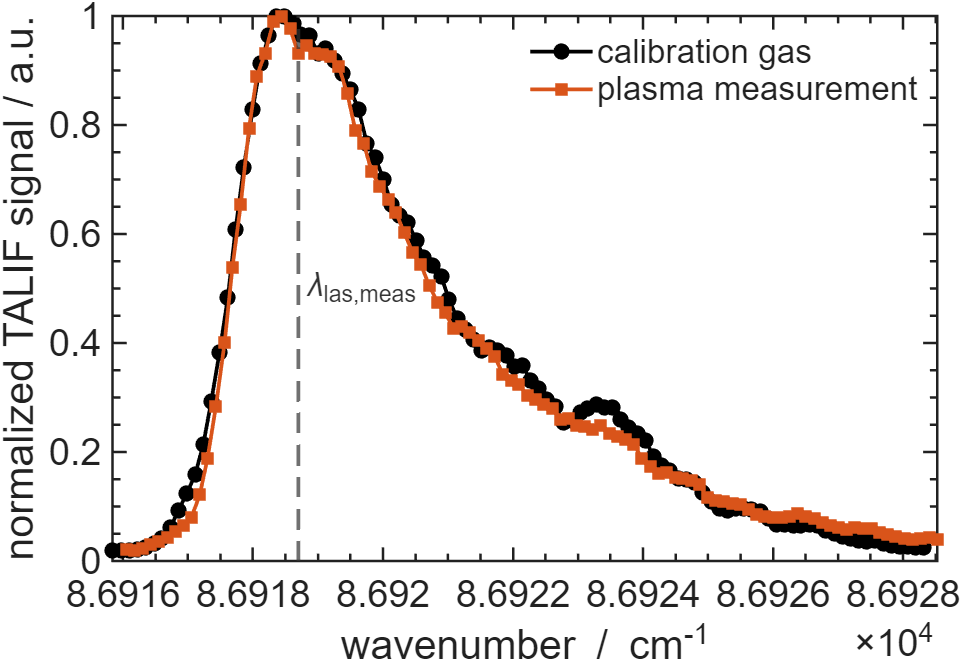}
    \caption{Wavelength TALIF excitation scan ($X^1\Sigma^+(v=0) \rightarrow B^1\Sigma^+(v'=0)$) with plasma switched on (orange, 700\,V voltage amplitude) and switched off (black) at $z=0$ (right below the plasma), with 70\,ppm CO.}
    \label{fig:signalWV}
\end{figure}

For this purpose, two excitation spectra were recorded using a calibration gas containing a known CO concentration (70\,ppm), as shown in Fig.~\ref{fig:signalWV}. In one case, the discharge was operated (orange curve), while in the other case, the discharge remained off (black curve), thereby avoiding additional gas heating. Measurements were conducted under conditions expected to produce the largest possible temperature difference, namely at the lowest gas flow (500\,sccm), highest applied voltage (700\,V amplitude), and at the position closest to the discharge.

As shown in Fig.~\ref{fig:signalWV}, both excitation spectra overlap almost perfectly. This indicates that no significant change in the rotational population distribution of the CO ground state occurs during plasma operation, at least in the region below the cavities where the CO density is measured in this study. Since the calibration gas is not additionally heated, a room-temperature rotational distribution can be assumed. The identical spectra therefore suggest that the measurements with the discharge switched on also correspond to approximately room temperature. In contrast to observations reported in \cite{damen_absolute_2019}, where elevated rotational temperatures led to noticeable spectral changes, the present atmospheric-pressure conditions provide sufficient collisional relaxation for the gas to equilibrate close to ambient temperature outside the discharge region.

Consequently, performing measurements at a single excitation wavenumber of approximately $86919~\mathrm{cm}^{-1}$ (corresponding to a single-photon wavelength of about 230.1\,nm) does not introduce significant systematic errors. Calibration measurements performed at the same wavelength remain valid, as the effective rotational temperature of CO is essentially unchanged. Moreover, at room temperature, the population of vibrationally excited levels in the electronic ground state of CO is negligible. The vibrational spacing of approximately 2170\,cm$^{-1}$ ($\approx 0.27$\,eV) \cite{le_floch_revised_1991} results in a Boltzmann population of the first excited vibrational level of only $N_1/N_0 \approx 3 \times 10^{-5}$ at $T=300$\,K. Hence, it is to be expected that essentially the entire CO population resides in the vibrational ground state ($v=0$).

\section{Basic diffusion model}
To interpret the experimental observations, a basic diffusion model analogous to that used for atomic oxygen in a previous study \cite{Steuer_2024} is applied. Unlike atomic oxygen in the previous model, which required the inclusion of chemical reactions due to its high reactivity, the generated CO molecules are chemically stable under the investigated conditions. Therefore, chemical reactions are neglected in the present model. The model is further extended to three dimensions, incorporates buoyancy-driven transport, and provides a more realistic representation of the gas flow within the reactor.

\begin{figure}[htb]
    \centering
    \includegraphics[width=\linewidth]{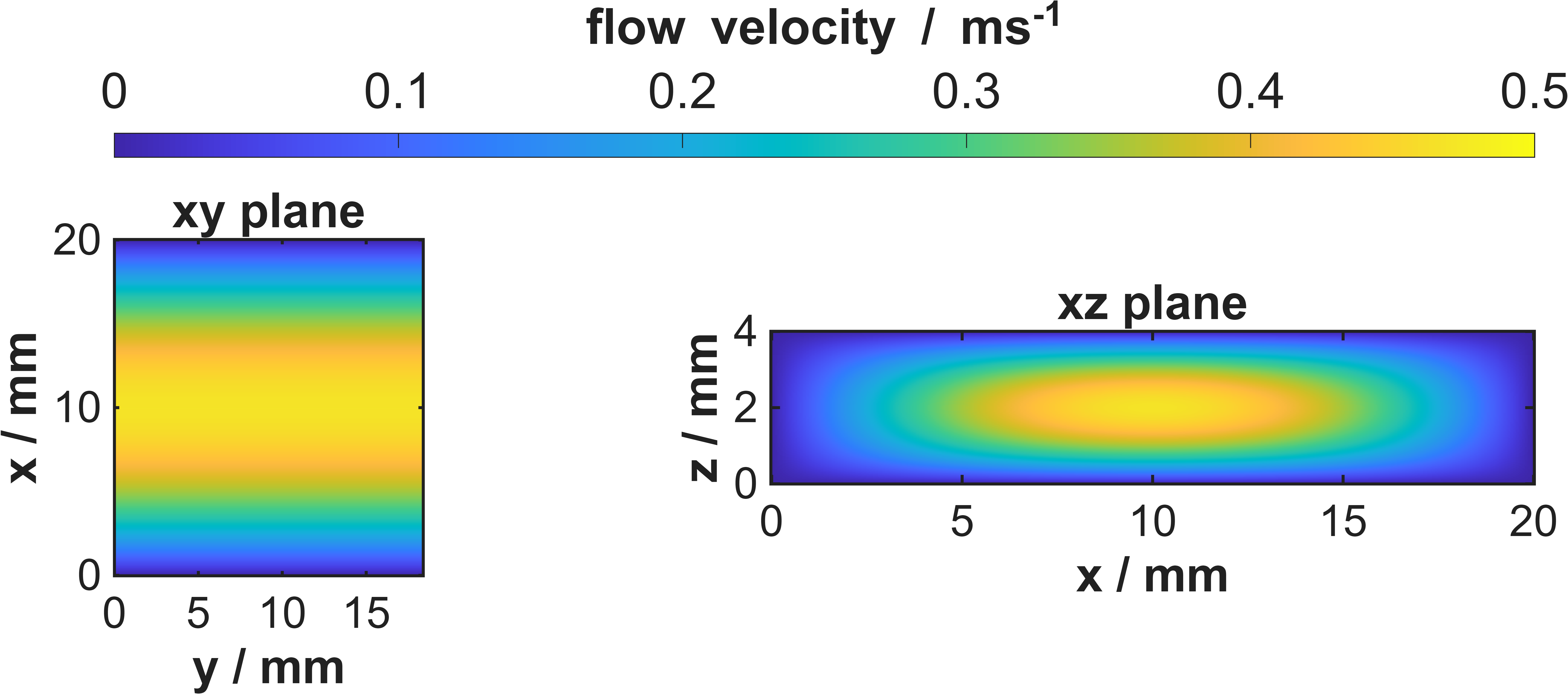}
    \caption{Poiseuille-like velocity profile for a total gas flow of 1\,slm.}
    \label{fig:flow}
\end{figure}

Diffusive transport is described by Fick's second law, which relates the temporal change in species concentration to the diffusion coefficient $D$. For carbon monoxide in helium at atmospheric pressure and room temperature, a diffusion coefficient of $D = 0.7\,\mathrm{m^2\,s^{-1}}$ is used \cite{fuller_new_1966}. \\
\begin{equation}
    \frac{\partial n}{\partial t} = D\Delta n
\end{equation}
When considering the stability of the molecules, it can be assumed that they are able to traverse the entire reactor volume. However, if a laminar gas flow in the $y$-direction is assumed, as in previous studies, this leads in practice to a spatially dependent velocity distribution $v_y(x,y,z)$ (higher velocities in the center and lower velocities near the walls), which should be taken into account. To this end, a Poiseuille-like velocity profile is assumed and scaled to match the total gas flow through the reactor. Figure \ref{fig:flow} shows the calculated velocity profile in the $xz$ and $yz$ plane for a total gas flow of 1\,slm.
In addition to the flow in the y-direction, buoyancy arising from the density difference between helium and CO leads to a non-negligible vertical transport in the $z$-direction. In practice, an equilibrium is established between buoyancy and viscous (Stokes) friction, resulting in an approximately constant velocity component along the $z$-direction. Since a detailed treatment of viscous friction is beyond the scope of this study, a constant vertical velocity of $v_z$ = 0.02 m/s was assumed, which is in good agreement with the experimental observations.

This allows a partial differential equation to be set up that takes diffusion and transport into account.
\begin{equation}
    \label{eq:model}
    \frac{\partial n}{\partial t} = D \left( \frac{\partial^2 n}{\partial x^2} + \frac{\partial^2 n}{\partial y^2} + \frac{\partial^2 n}{\partial z^2}\right) - v_y(x,y,z) \frac{\partial n}{\partial y} - v_z \frac{\partial n}{\partial z}
\end{equation}
This partial differential equation was solved using an explicit finite-difference scheme. A prescribed concentration, such as a measured value, was applied as the boundary condition at $z$ = 0\,mm, corresponding to the interface between the cavities and the region below.

All other boundaries were treated as follows: reflective (zero-flux, Neumann) conditions were applied at the walls in the $x$-direction and at the top boundary in the $z$-direction. In the $y$-direction, the inflow at $y = 0$ was set to zero CO concentration, while the outflow at $y = 18.4$\,mm was treated with an upwind scheme to allow CO to leave the domain consistently with the local flow.

A grid with 500 points in the $y$-direction (covering 18.4\,mm) is used. In the $z$-direction, 40 points cover 4\,mm and in the $x$-direction, 50 points cover 20\,mm. About 30000 time steps of 1\,µs are needed for convergence.

Although several physical and chemical effects are neglected, the basic diffusion model reproduces the measured density distributions reasonably well. This suggests that the model captures the dominant transport processes in the region below the cavities. However, in a more advanced model, the boundary conditions would require further refinement. For example, the cavities and the metal grid not only generate CO but also act as sinks through surface adsorption and recombination processes, which are not included in the present approach. Furthermore, a spatially uniform temperature is assumed, although significant temperature gradients are expected between the cavities and the surrounding volume \cite{Steuer_2023,vanImpel_2025}. In addition, gas-phase chemical reactions in the volume, such as reactions involving reactive oxygen species, are entirely neglected. However, including these effects would significantly complicate the model and exceed the scope of this work.\\

\section{Results}
The results are presented as follows: First, a three-dimensional map obtained under standard conditions (1\,slm He + 4\,sccm CO$_2$) is shown to illustrate the measurement routine and the capabilities of the diagnostic setup. In addition, a direct comparison with the diffusion model is provided. To further investigate the influence of the gas flow on the CO distribution and to verify the assumed laminar flow pattern, two-dimensional data from a height scan ($z$-scan) are presented. Finally, the influence of the applied voltage amplitude and the CO$_2$ admixture is examined using spatially averaged (0D) data in order to gain further insight into the discharge dynamics and the overall dissociative potential of the MCPA reactor.

\subsection{3D map}
\begin{figure*}[h]
    \centering
    \includegraphics[width=\textwidth]{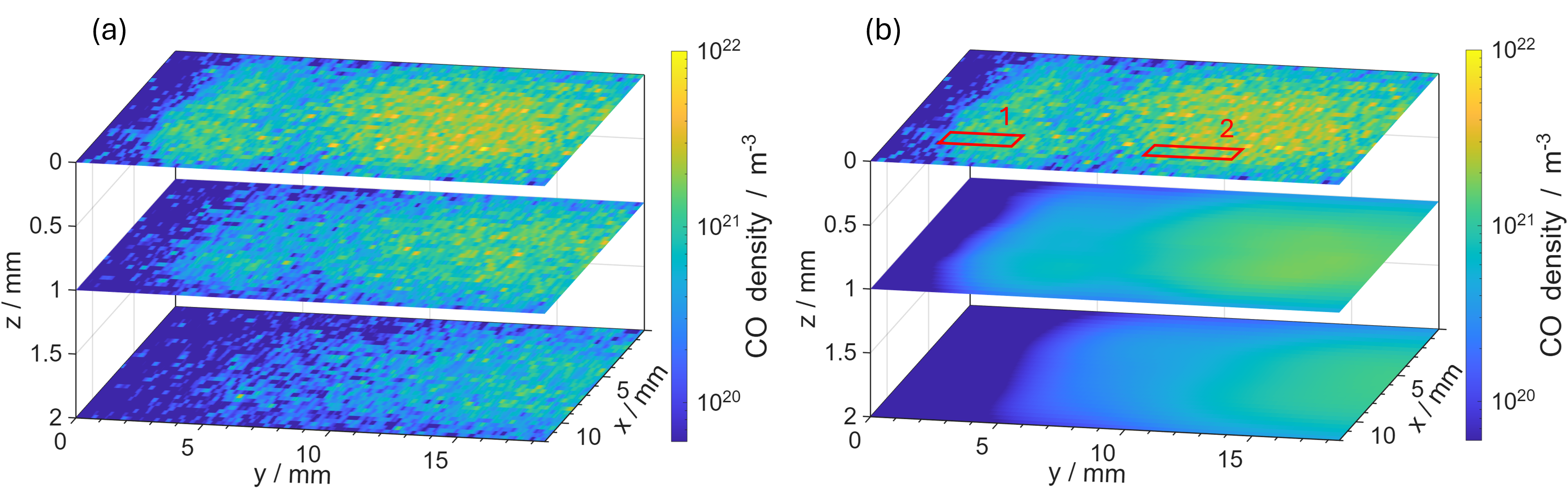}
    \caption{(a) Measured three-dimensional CO density distribution where $z=0$ corresponds to the measurement plane directly below the cavities. Experimental conditions: 1\,slm He with 0.4\,\% CO$_2$ admixture; applied voltage amplitude 600\,V; $x$-step width = 400\,µm; ICCD gate time = 100\,ns; exposure time per position = 60\,s. (b) Comparison with the diffusion model results using the measured $z=0$ map as input. The regions of interest (ROIs, \textit{position~1} \& \textit{2}) used for the analysis in Chapter~\ref{chap:VT} are indicated for the $z=0$ plane.}
    \label{fig:3D_scan}
\end{figure*}

Figure~\ref{fig:3D_scan} shows the reconstructed three-dimensional CO density distribution within the MCPA reactor for an admixture of 4\,sccm CO$_2$ in a helium flow of 1\,slm. The dataset is represented by three two-dimensional heatmaps corresponding to the field of view (FOV) indicated in Fig.~\ref{fig:TALIF_setup}(b), separated by 1\,mm in the $z$-direction. The position $z=0$ corresponds to the measurement plane closest to the reactor surface without partial clipping of the laser beam ($\approx$100\,µm above the nickel grid).

At this position, the spatial structure directly below the cavities becomes clearly visible. Since not all cavities ignite under these operating conditions, localized regions of enhanced CO density are observed. The measured structures agree well with the discharge photograph shown in Fig.~\ref{fig:TALIF_setup}(a), confirming that CO production occurs predominantly below the actively ignited cavities. The gas flow direction is from left to right, and two main active regions can be identified in the $y$-direction between approximately 2--6\,mm and again starting from about 8\,mm. This agreement demonstrates that the spatial resolution of the CO-TALIF measurements is sufficient to reproduce the discharge patterns observed optically.

The absolute CO number density reaches values on the order of $10^{21}\,\mathrm{m^{-3}}$, with local maxima approaching $10^{22}\,\mathrm{m^{-3}}$, corresponding to a dissociation degree of approximately 10\,\% directly outside the discharge region. However, individual cavities cannot be fully resolved spatially, as the detection integrates over the laser beam width of approximately 180\,µm in the $x$-direction, which is comparable to the cavity diameter. In addition, rapid diffusion of CO into neighboring regions smooths local gradients, preventing a clear distinction between active discharge zones (i.e., cavities) and areas consisting solely of the nickel surface. Furthermore, the laser center must remain at least 95\,µm away from the surface to avoid beam clipping, which imposes an additional limitation on the achievable spatial resolution close to the electrode.

An accumulation effect along the array structure is also observed, as each successive discharge region contributes additional converted CO$_2$ to the overall CO density along the flow direction. This behavior will be discussed further in the context of voltage and admixture variations.

With increasing distance from the cavities, the CO density decreases due to diffusion and mixing with incoming fresh gas, resulting in strong downstream dilution. Consequently, the structures observed at $z=0$ are progressively displaced in the flow direction. This behavior becomes more pronounced at larger distances from the cavities and is consistent with a Poiseuille-like flow profile between the nickel foil and the lower quartz plate of the reactor cover. The flow velocity reaches its maximum near the channel center ($z \approx 2\,\mathrm{mm}$), leading to enhanced convective transport compared to regions closer to the surfaces.

This diffusion-driven behavior can be described by the simplified transport model shown in Fig.~\ref{fig:3D_scan}(b). The model uses the measured two-dimensional CO density distribution at $z=0$ as an input and calculates the corresponding distributions at $z=1$\,mm and $z=2$\,mm by accounting for diffusion and convective flow.

A direct comparison between experiment and simulation reveals very good agreement in both the spatial distribution and the downstream displacement of the density structures. In particular, the flow-induced drift of the cavity-related structures is clearly reproduced by the model. This agreement indicates that no additional dominant transport or reaction processes need to be considered and that the observed dynamics are primarily governed by diffusion and convective flow.

To further validate the transport description and assess the model under varying conditions, the gas flow was systematically varied. The influence of different flow rates on the CO distribution and the applicability of the model are discussed in the following section.

\subsection{Flow variation}
\begin{figure*}[htb]
    \centering
    \includegraphics[width=\textwidth]{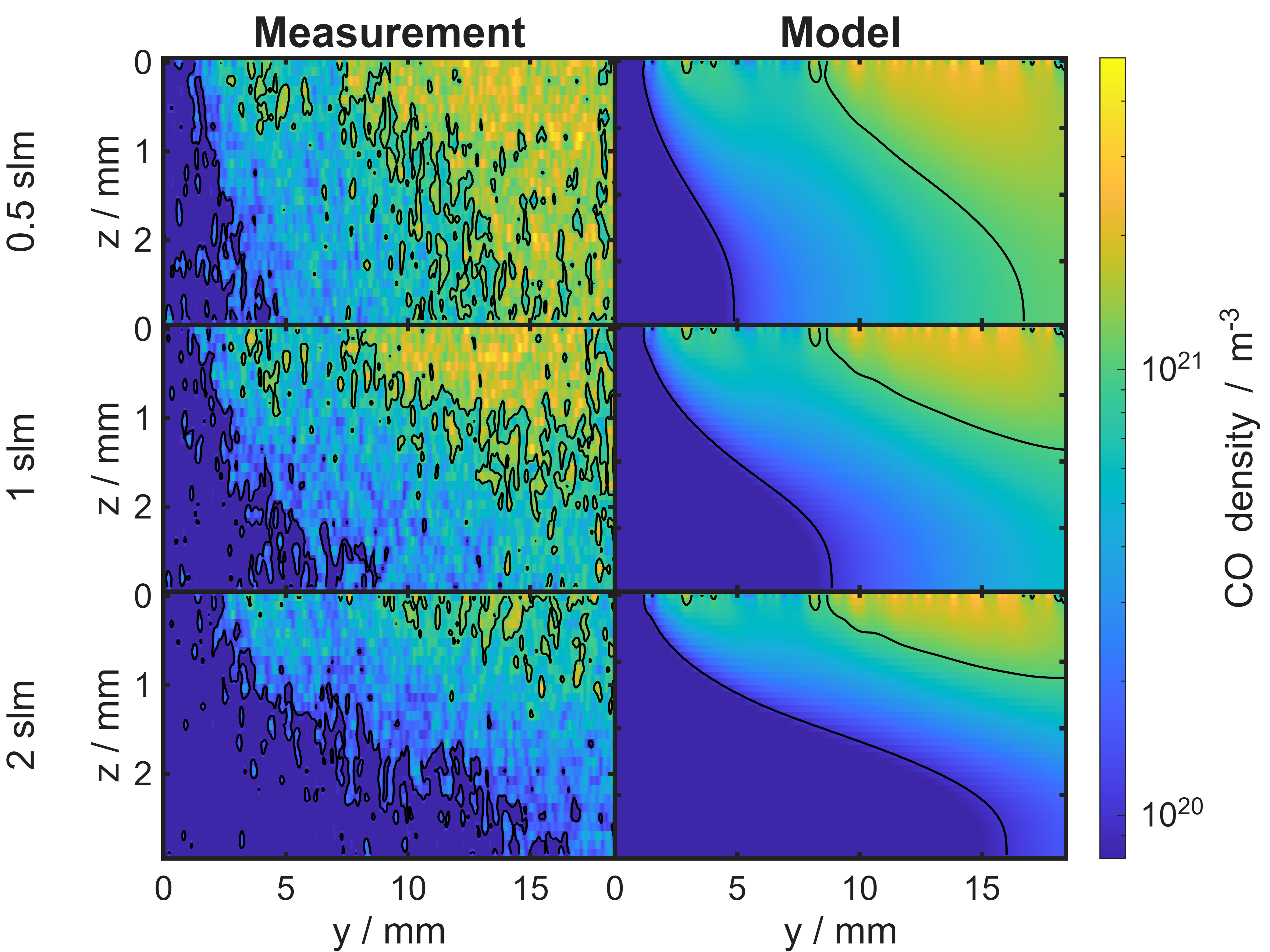}
    \caption{Measured (left) and modeled (right) height scan of the CO density for different applied flows 0.5--2\,slm at a fixed x-position ($x=6.8$\,mm). Conditions: 99.6\,\% He + 0.4\,\% CO$_2$; 600\,V amplitude; step width = 100\,µm; gate time = 100\,ns; exposure time per position = 60\,s.}
    \label{fig:TALIF_zscans}
\end{figure*}

In Fig.~\ref{fig:TALIF_zscans}, CO density profiles obtained for different applied gas flows are presented. Here, the spatial resolution of the measurement in the $z$-direction was increased by changing the step width to 100\,µm. The measurements shown on the left-hand side correspond to a fixed lateral position close to the center ($x=6.8\,\mathrm{mm}$, Fig.~\ref{fig:3D_scan}). For all investigated flow conditions, a constant admixture of 0.4\,\% CO$_2$ in helium was maintained.

The admixture was intentionally kept constant so that both the number and spatial distribution of ignited cavities remained unchanged, thereby eliminating variations in discharge activity as a confounding factor. This allows the observed changes in CO density to be primarily attributed to modified transport conditions rather than altered discharge behavior.

For the corresponding three-dimensional simulations, the same measured 2D CO density distribution at $z=0$ as used in Fig.~\ref{fig:TALIF_zscans} served as the model input. Only the flow velocity was adjusted according to the respective experimental conditions. This approach enables a direct assessment of the influence of convective transport on the measured density profiles while keeping the CO production term constant.

A comparison between measurement and simulation again shows very good agreement over the full range of applied flow rates. In particular, the model consistently reproduces the progressive downstream distortion of the CO density profile with increasing flow velocity. The effect is most pronounced near the channel center (z=2\,mm), where the highest flow velocities are expected.

In addition, an accumulation of CO can be observed in front of the glass cover located at $z=4$\,mm. This effect becomes particularly evident when following an iso-density contour line (black line), which clearly visualizes the redistribution of the gas. The contour shape is well reproduced by the simulation and reflects the underlying flow profile within the reactor channel.

This agreement supports the validity of the applied diffusion coefficient and the assumed flow profile. The observed behavior is consistent with predominantly laminar flow conditions within the reactor, with no indications of strong turbulence or vortex-like structures such as those reported, for example, in \cite{boddecker_interactions_2023} for a comparable SDBD reactor. While a more detailed fluid-dynamic analysis would be required for a rigorous verification, the present results indicate that the adopted transport description provides a physically consistent approximation.

\subsection{Voltage $\mathrm{\&}$ admixture variation}\label{chap:VT}

Having established that the transport dynamics in the reactor are dominated by laminar flow and diffusion, the CO density distribution directly below the cavities can be considered largely unaffected by complex flow-induced mixing. This allows a more direct investigation of the discharge characteristics themselves. In the following, the influence of the applied voltage amplitude and the CO$_2$ admixture on the CO production is therefore examined.

    \begin{figure}
    \centering
    \includegraphics[width=\linewidth]{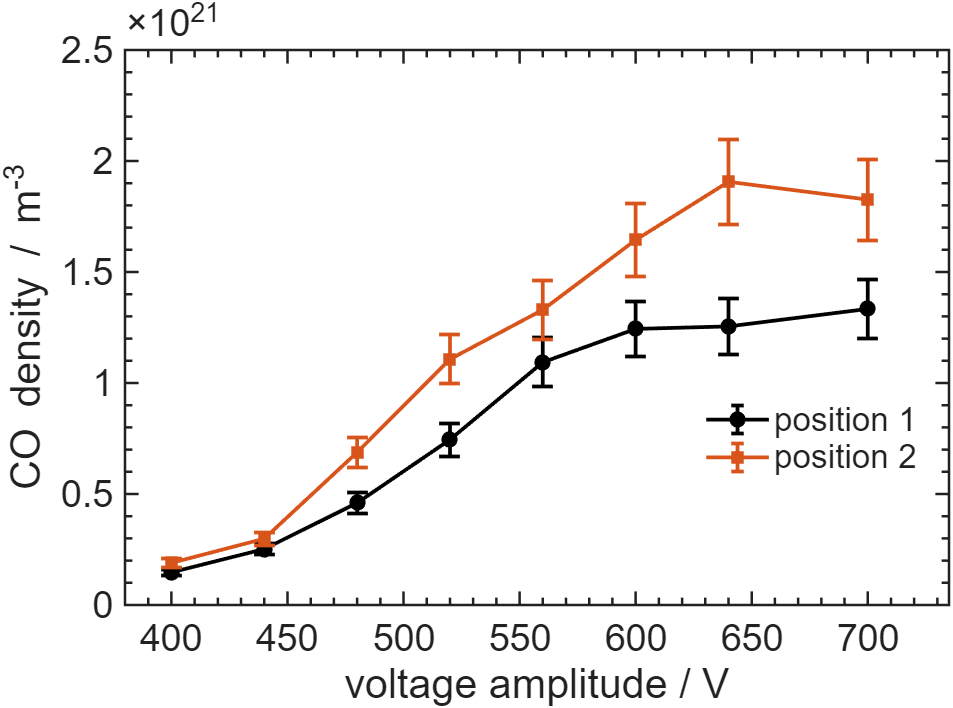}
    \caption{CO density against the applied voltage amplitude for an averaged area at the first igniting cavities (position 1) and a position downstream (position 2). Conditions: 1\,slm He + 4\,sccm CO$_2$; gate time = 100\,ns; exposure time per position = 60\,s.}
    \label{fig:voltage}
\end{figure}

To further investigate the discharge characteristics, the applied voltage was systematically varied. In order to reduce the multidimensional datasets to representative 0D quantities suitable for comparison, measurements were evaluated at a fixed $z$-position and spatially averaged over defined rectangular regions in the $xy$-plane.

The selected height was $z=0$, i.e., directly below the cavities, where dilution effects due to mixing with the incoming fresh gas are expected to be minimal. The first rectangular region of interest (ROI), indicated in Fig.~\ref{fig:3D_scan}(b) and referred to as \textit{position~1}, was chosen to cover an area where the majority of cavities are ignited, thereby minimizing variations caused by additional ignition events within the ROI. Furthermore, the region was positioned to include the first igniting cavities in flow direction, avoiding artificial accumulation effects from downstream discharge regions. The spatial coordinates of \textit{position~1} are $y=2.7$--$5.4$\,mm and $x=9.2$--$10.4$\,mm (in 300\,µm steps), including about 27 cavities.

To investigate the influence of downstream transport and accumulation effects, a second rectangular ROI, denoted as \textit{position~2}, was defined further along the flow direction at $y=10.8$--$14.4$\,mm, while keeping the same $x$- and $z$-positions (about 36 cavities).

Figure~\ref{fig:voltage} shows the variation of the CO number density for applied voltages between 400 and 700\,V. At both evaluated positions, the density increases from approximately $0.2 \cdot 10^{21}\,\mathrm{m^{-3}}$ to about $1.3 \cdot 10^{21}\,\mathrm{m^{-3}}$ at \textit{position~1} and up to $1.9 \cdot 10^{21}\,\mathrm{m^{-3}}$ at \textit{position~2}, respectively.

For lower voltages, the increase is gradual and becomes steeper with increasing voltage. At \textit{position~1}, a tendency toward saturation is observed starting at approximately 560\,V, whereas at \textit{position~2} a similar behavior appears at slightly higher voltages around 600\,V. Over the entire voltage range, the CO densities measured at \textit{position~2} exceed those at \textit{position~1}, and the difference between both positions becomes more pronounced at higher voltages.

The initial increase in CO density at lower voltages can be attributed to enhanced power input into the discharge, leading to increased electron impact dissociation of CO$_2$. In addition, the effective discharge area expands with increasing voltage due to a growing number of ignited cavities within the observed region. This results in a higher spatially averaged CO production rate and consequently higher measured densities.

The increasing separation between \textit{position~1} and \textit{position~2} with rising voltage can be explained by cumulative production effects along the flow direction. With a larger number of ignited cavities between the two regions at higher voltages, additional CO is produced upstream of \textit{position~2}, contributing to the convective transport and leading to higher downstream densities.

The observed tendency toward saturation indicates that the local dissociation within the cavities approaches a high conversion regime. Similar behavior has previously been reported for the molecular oxygen model system, where nearly complete dissociation inside individual cavities was observed and saturation effects occurred \cite{Steuer_2023,vanImpel_2025}. The dissociation energy of CO$_2$ is approximately 5.45\,eV \cite{gong_bond_2024}, while the dissociation energy of O$_2$ is about 5.1\,eV \cite{brix_fine_1954}. Since these energies are of similar magnitude, it can be assumed as a first approximation that both molecules exhibit comparable dissociation behavior in the discharge.

However, the maximum spatially averaged CO densities measured in the present study correspond only to apparent dissociation degrees of approximately 1.4\,\% at \textit{position~1} and 1.9\,\% at \textit{position~2}. The apparent contradiction between a low overall conversion and a high local dissociation inside the cavities can be solved when geometrical and temporal dilution effects are taken into account.

The cavities have an edge-to-edge spacing of 0.15\,mm and a diameter of 0.2\,mm, resulting in a cavity area of approximately 0.031\,mm$^2$ and a surrounding nickel surface area of about 0.092\,mm$^2$. Thus, the active discharge area represents only roughly one-third of the total surface area within one unit cell. In addition, under admixture of molecular gases the cavity is typically not fully filled by the discharge; a filling fraction of approximately 25\,\% has been reported \cite{Steuer_2025}. This introduces an additional geometrical reduction factor of about four.

Furthermore, the effective plasma-on time is limited. For a driving frequency of 15\,kHz (period 66.7\,µs), the discharge duration is approximately 15\,µs per ignited half-cycle in the oxygen case. Since the plasma is active only during one half-phase, the effective duty cycle corresponds to roughly a factor of two reduction when averaged over a full period.

Combining these geometrical and temporal factors yields an overall dilution factor of approximately 
\[
3 \times 4 \times 2 \approx 24.
\]
Taking into account additional transport limitations, such as incomplete escape of CO molecules from the cavity before recombination or back-reaction occurs, the effective local dissociation degree within the active plasma region is expected to be significantly higher than the spatially averaged value suggests. 

Under these considerations, the experimentally observed saturation behavior at both positions is consistent with a locally high dissociation degree inside the cavities and reaching an equilibrium between dissociation and back reactions to CO$_2$.     

    \begin{figure}
    \centering
    \includegraphics[width=\linewidth]{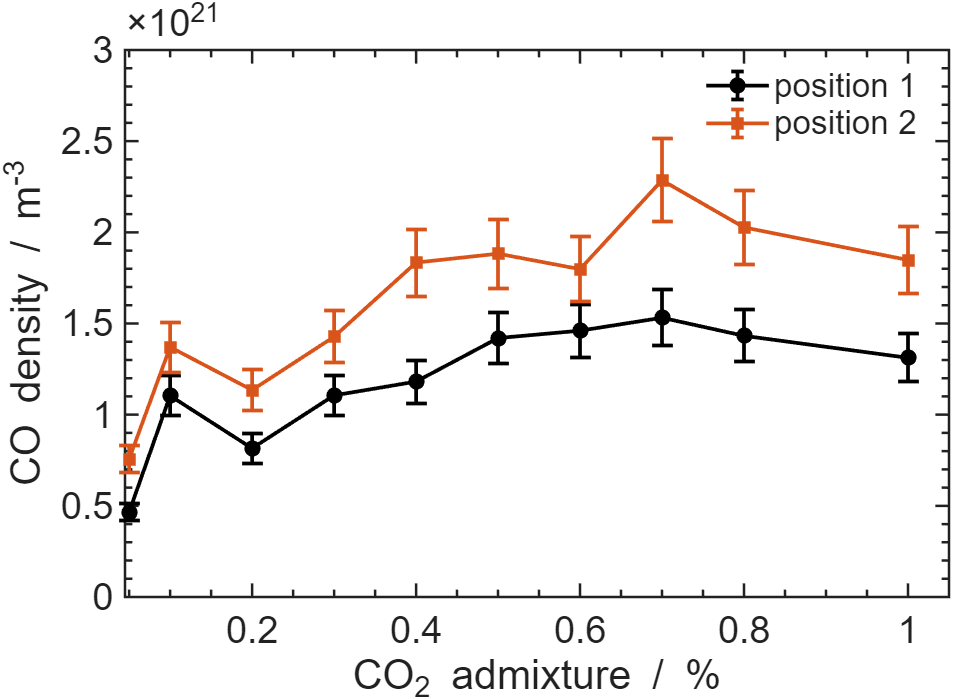}
    \caption{CO density against the admixed CO$_2$ concentration for an averaged area at the first igniting cavities (position 1) and a position downstream (position 2). Conditions: 1\,slm He + x\,\% CO$_2$; 600\,V amplitude; gate time = 100\,ns; exposure time per position = 60\,s.}
    \label{fig:admix}
\end{figure}

The variation of the CO$_2$ admixture is shown in Fig.~\ref{fig:admix} for concentrations between 0.05\,\% and 1\,\% at a fixed applied voltage amplitude of 600\,V. The measured CO density increases overall with increasing admixture, aside from a small drop at 0.2\,\%. A maximum is reached at approximately 0.7\,\% CO$_2$, after which the density decreases again. Both observation positions show the same qualitative behavior, although the difference between the two densities increases with increasing admixture. The maximum CO densities reached are about $1.5\cdot10^{21}$\,m$^{-3}$ at \textit{position~1} and $2.3\cdot10^{21}$\,m$^{-3}$ at \textit{position~2}.

The initial increase in CO density is mainly caused by the larger amount of admixed CO$_2$. As discussed previously, the dissociation degree within the cavities is expected to be high; therefore, increasing the CO$_2$ concentration directly increases the amount of CO that can be produced. This behavior is particularly visible in the first two measurement points, where doubling the CO$_2$ admixture approximately doubles the measured CO density.

However, this linear trend does not persist for higher admixtures. After the next increase in CO$_2$ concentration, the CO density initially decreases. One possible explanation is a change in the discharge behavior inside the cavities. Similar effects have been observed for oxygen admixtures, where the cavity filling of the discharge changed significantly with increasing molecular gas content \cite{Steuer_2025}. The observed drop could therefore indicate a possible discharge mode transition from a more glow-like discharge to a filamentary streamer-based discharge with shorter ignition per cycle, although further investigations are required to confirm this interpretation.

For higher CO$_2$ concentrations, the CO density increases again but with a noticeably reduced slope. This suggests that the available discharge power is no longer sufficient to maintain the same effective dissociation degree, as the CO$_2$ dissociation process requires substantial energy input.

The decrease in CO density observed beyond an admixture of about 0.7\,\% is attributed to a reduction in the number of ignited cavities within the ROI. As fewer cavities remain active, the number of CO production sites decreases, which reduces the spatially averaged density. While the CO production per individual cavity may still increase, the overall production is limited by the reduced number of active discharges. In addition, a shorter ignition duration at higher molecular admixtures could further decrease the effective production time.

\section{Conclusion}
Carbon monoxide densities above a micro-cavity plasma array (MCPA) were successfully measured using CO-TALIF. The applied diagnostic setup, combining TALIF with an ICCD camera, enabled spatially resolved measurements and allowed the reconstruction of three-dimensional CO density distributions within the reactor. The measured density structures clearly reproduced the spatial pattern of ignited cavities, demonstrating that the achieved spatial resolution is sufficient to resolve structural variations within the discharge array.

The experimentally observed CO distributions could be reproduced well by a simplified three-dimensional diffusion model. This agreement indicates that the transport of CO below the cavities is mainly governed by diffusion and convective gas flow, without the need to introduce additional complex processes.

The combined experimental and modeling results from the flow variation further show that the gas flow inside the reactor follows a laminar, Poiseuille-like profile. No indications were found that the plasma discharge introduces significant turbulence into the gas phase. In addition, the observed transport behavior is consistent with literature values for the diffusion coefficient of CO in helium, providing further validation of the applied model.

The voltage and admixture variations provide insight into the plasma chemistry inside the cavities. The observed saturation behavior in the voltage dependence suggests that very high dissociation degrees are reached within the microdischarges. This indicates that the MCPA geometry is particularly efficient for dissociation-driven processes.

Overall, the combination of an MCPA reactor with spatially resolved CO-TALIF diagnostics provides a powerful tool for studying plasma-assisted conversion processes. The ability to resolve spatial structures within the discharge array opens the possibility to systematically investigate plasma–catalyst interactions, for example by introducing different catalysts or loadings of catalysts into selected cavities. Combined with complementary diagnostics for surface charges, electric fields, and atomic oxygen measurements, this approach provides a robust platform for future studies of plasma-catalyst interactions, including the conversion of carbon-based molecules.
\ack
This work is funded by the Deutsche Forschungsgemeinschaft (DFG, German Research Foundation) via SFB 1316 (project A6).

\section{References}
\bibliographystyle{iopart-num}
\bibliography{references.bib,CO-Measurement.bib,Motivation_CO.bib,Schlieren-PIV.bib}
\end{document}